\documentclass[a4paper,12pt,tightenlines,notitlepage,nofootinbib]{revtex4-2}
\usepackage{amsmath,amssymb,setspace}
\usepackage{graphicx}
\usepackage{verbatim}
\usepackage{newtxtext}
\usepackage{newtxmath}
\usepackage[normalem]{ulem}
\usepackage[nodayofweek,ddmmyy]{datetime}
\usepackage[x11names]{xcolor}
\usepackage[unicode,colorlinks,citecolor=Blue3,linkcolor=Blue3,bookmarks=true]{hyperref}

\newcommand{\eg}{{\it e.g.\,}}
\newcommand{\const}{\mathop{\rm const}\nolimits}
\newcommand{\svector}[2]{\begin{pmatrix}#1 \\ #2 \end{pmatrix}}

\DeclareSymbolFont{bbold}{U}{bbold}{m}{n}
\DeclareSymbolFontAlphabet{\mathbbold}{bbold}

\begin{document}



\title{Increasing quantum speed meter sensitivity using optical spring}


\author{L.~A.~Barinov}
\affiliation{Russian Quantum Center, Skolkovo IC, Bolshoy Bulvar 30, bld.\ 1, Moscow, 121205, Russia}
\affiliation{Moscow Institute of Physics and Technology, 141700 Dolgoprudny, Russia}
\author{F.~Ya.~Khalili}
\email{farit.khalili@gmail.com}
\affiliation{Russian Quantum Center, Skolkovo IC, Bolshoy Bulvar 30, bld.\ 1, Moscow, 121205, Russia}

\begin{abstract}

The double-pass interferometer scheme was proposed in Ref.\,[Light Sci. Appl. {\bf 7}, 11 (2018)] as the method of implementation of the quantum speed meter concept in future laser gravitational-wave (GW) detectors. Later it was shown in Ref.\,[Phys. Rev. D {\bf 110}, 062006 (2024)] that it allows to implement the new type of the optical spring that does not require detuning of the interferometer. Here we show that both these regimes can coexist, combining the speed meter type broadband sensitivity gain with the additional lows-frequency minimum in the quantum noise originated from the optical spring. We show that the location of this minimum can be varied without affecting the core optics of the interferometer, allowing to tune the quantum noise shape in real time to follow the ``chirp'' GW signals.

\end{abstract}

\maketitle


\section{Introduction}

Sensitivity of the best modern optomechanical sensors is limited by the constraints imposed by quantum mechanics. In particular, sensitivity of the contemporary laser interferometric gravitational-wave (GW) detectors \cite{CQG_32_7_074001_2015, Acernese_CQG_32_024001_2015, Tse_PRL_123_231107_2019_short, Acernese_PRL_123_231108_2019_short, Ganapathy_PRX_13_041021_2023, Jia_Science_385_1318_2024} is close to the Standard Quantum Limit (SQL) that corresponds to the balance of measurement imprecision and the quantum back action of the meter originating from the Heisenberg uncertainty relation \cite{67a1eBr, 74a1eBrVo, Caves_RMP_52_341_1980, 92BookBrKh}.

Several methods for bypassing the SQL that can be used in GW detectors have been proposed, see the reviews \cite{12a1DaKh, 19a1DaKhMi}. In particular,  it was shown by W.~Unruh in Ref.\,\cite{Unruh1982} that injecting squeezed light with the optimally tuned frequency-dependent squeeze angle into the interferometer, it is possible to suppress the quantum noise spectral density proportionally to the squeeze factor of the incident light $e^{2r}$. A practical method for generating frequency-dependent squeezed light was later proposed  in Ref.\,\cite{02a1KiLeMaThVy}. It was shown in that work that the necessary frequency dependence can be created by reflecting a frequency-independent squeezed vacuum from an additional so-called filter cavity.

The disadvantage of this scheme is its vulnerability to optical losses in the filter cavity. These losses can be mitigated using very long filter cavities, with lengths comparable to the length of the main interferometer arms. However, this approach requires significant modification of the existing interferometers infrastructure. Recently, relatively short 300-meter filter cavities fitting into the existing infrastructure were added to the two LIGO GW detectors \cite{Ganapathy_PRX_13_041021_2023}, providing the modest (about 3 dB) reduction of the quantum noise in the frequency band of about one octave \cite{Jia_Science_385_1318_2024}.

Another more radical method of overcoming the SQL, known as the quantum speed meter,  was formulated in Ref.\,\cite{90a1BrKh}. The basic idea of this method is the measurement of the velocity of the probe object instead of its position. Several implementations of this concept suitable for the laser GW detectors were proposed, see Refs.\,\cite{00a1BrGoKhTh, Purdue2002, Chen2002, 04a1Da, 17a1KnDaHiKh,  18a1DaKnVoKhGrStHeHi, Huttner_CQG_34_024001_2017}. The detailed discussion of these proposals can be found in Sec.\,5 of the review \cite{19a1DaKhMi}.

\begin{figure}
  \includegraphics[scale=1.2]{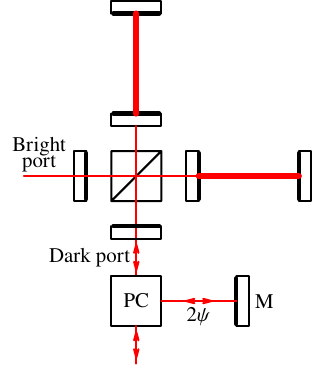}
  \caption{The two-pass interferometer. PC: polarization circulator; M: mirror.}\label{fig:scheme}
\end{figure}

The double-pass scheme proposed in Ref.\,\cite{18a1DaKnVoKhGrStHeHi} is, in our opinion, the most interesting because it maintains the core dual-recycled topology of contemporary GW detectors (see \eg Ref.\,\cite{CQG_32_7_074001_2015}) unchanged, with the only new requirement that the interferometer has to be able to work with two orthogonal polarization of the light. This scheme is shown in Fig.\,\ref{fig:scheme}. Here both polarizations of the carrier light in the interferometer are equally excited through the bright port. In the dark port, an additional polarization circulator ("PC") is located, which modifies the polarization of the light passing through it and reroutes the light in the following way. (i) The incident beam --- just vacuum noise or the squeezed vacuum --- goes directly to the interferometer. (ii) The returning beam is redirected to the additional mirror M. (iii) The reflected beam goes to the interferometer again. (iv) Finally, the second returning beam exits the interferometer and is measured by the homodyne detector.
During the steps (ii) and (iii), an additional phase shift $2\psi$ is introduced into the light. The possible design of the PC block is shown in Fig.\,3 of Ref.\,\cite{18a1DaKnVoKhGrStHeHi}.

The value of $2\psi=\pi$ inverts the probing beam sign, implementing thus the original double-measurement concept of the quantum speed meter shown in Fig.\,1 of Ref.\,\cite{90a1BrKh}. At the same time, it was shown in Ref.\,\cite{24a1Kh} that other values of $\psi$ allow to increase sensitivity in another way. If $\sin2\psi\ne0$, then the optical spring effect arises in this scheme, effectively converting the near free-mass mechanical degree of freedom of the scheme into the harmonic oscillator and thus reducing the SQL in the frequency band around the mechanical resonance frequency \cite{12a1DaKh}. It is worth noting here that opposite to the ``canonical'' \cite{99a1BrKh, 01a1BrKhVo, 01a2Kh, Buonanno2002, 12a1DaKh} type of the optical spring, this ``double-pass'' one does not require the detuning of the interferometer and therefore could be more convenient for implementation in the GW detectors.

It is natural to expect that by augmenting the quantum speed meter with the optical spring, additional sensitivity gain can be achieved. In this paper, we analyze this combined regime in detail and show that indeed the optical spring allow to increase the quantum speed meter sensitivity in the frequency band around the mechanical eigenfrequency $\Omega_0$.

In order to avoid overcluttering of the equations in this ``proof of principle'' paper, we do not take into account explicitly the optical losses here. Instead, we rely on the work \cite{18a1DaKnVoKhGrStHeHi}, where the thorough analysis of the losses was done for the parameters values close to ones used here. The key difference of the scheme considered here form that of \cite{18a1DaKnVoKhGrStHeHi} is the additional low-frequency sensitivity gain provided by the optical spring, and it is known that this kind of the sensitivity gain is robust to the losses, see Ref.\,\cite{12a1DaKh}.

The paper is organized as follows. In the next section, we introduce the main notations and assumption used throughout the paper and remind the reader the basic features of the position meter and speed meter quantum noises. 
Then in Sec.\,\ref{sec:S} we derive the quantum noise spectral density of our scheme, find the convenient asymptotic approximations for this spectral density, and provide the sensitivity estimates for our scheme. Finally, in Sec.\,\ref{sec:conclusion}, we summarize the results of this work.

\section{Quantum noises of the position meter and the speed meter}\label{sec:pm_vs_sm}

We quantify the quantum noise by means of the following convenient dimensionless factor:
\begin{equation}\label{xi2}
  \xi^2(\Omega) = \frac{S_{\rm sum}(\Omega)}{S_{\rm SQL}(\Omega)} \,,
\end{equation}
where $\Omega$ is the running frequency,
\begin{equation}\label{S_gen}
  S_{\rm sum}(\Omega) = |\chi^{-1}(\Omega)|^2S_{xx}(\Omega)
    + 2\Re[\chi^{-1}(\Omega)S_{xF}(\Omega)] + S_{FF}(\Omega) \,,
\end{equation}
is spectral density of the force-normalized (more convenient one for the non-free-mass probe objects, in comparison with the position-normalized one) sum quantum noise, $S_{xx}$ and $S_{FF}$ are spectral densities of, respectively, the measurement imprecision noise and the back action force, $S_{xF}$ is the cross-correlation spectral density of these noises,
\begin{equation}\label{S_SQL}
  S_{\rm SQL}(\Omega) = \hbar m\Omega^2
\end{equation}
is the free mass SQL, $m$ is the probe mass, and $\chi^{-1}$ is its response function (see details in the review \cite{12a1DaKh}). For example, for a free mass, $\chi^{-1}=-m\Omega^2$, and for a harmonic oscillator, $\chi^{-1}=K-m\Omega^2$, where $K$ is the spring constant. In Eq.\,\eqref{S_SQL}, as well as throughout this paper, we use the double-sided  normalization of the spectral densities with the frequency $\Omega$ varying from $-\infty$ to $\infty$.

The component spectral denstities that appear in the R.H.S. of Eq.\,\eqref{S_gen} satisfy the following uncertainty relation, see Refs.\,\cite{92BookBrKh, 20a1KhZe}:
\begin{equation}\label{S_xS_F}
  S_{xx}(\Omega)S_{FF}(\Omega) - |S_{xF}(\Omega)|^2 \ge \frac{\hbar^2}{4} \,.
\end{equation}
In this paper, we neglect the optical losses, which corresponds to the exact equality in this equation.

We assume that the phase (sine) quadrature of the probing light  $\hat{a}_s$ is squeezed and the amplitude (cosine) one  $\hat{a}_c$ is proportionally anti-squeezed. Correspondingly, their spectral densities are equal to
\begin{equation}\label{S_SQZ}
  S[\hat{a}_s] = \frac{e^{-2r}}{2} \,, \quad S[\hat{a}_c] = \frac{e^{2r}}{2} \,.
\end{equation}

\begin{table}
  \begin{ruledtabular}
    \begin{tabular}{llll}
      Description & Notation & Value \\
      \hline
      Arm length & $L$ & 4 km \\
      Test-mass & $m$ & 100 kg \\
      Circulating power & $I_c$ & $2\times1.5\,{\rm MW}$ \\
      Squeeze factor & $e^{2r}$ & 4 (6 dB) \\
      Normalized optical power & $\Theta_{\rm SM}=2\Theta_{\rm PM}$ &
        $7\times10^8\,{\rm s}^{-3}$
    \end{tabular}
  \end{ruledtabular}
  \caption{Parameter values, used for the estimates. Mostly, they corresponds to the planned future iteration ${\rm A}^\#$ of the LIGO GW detector, see Ref.\,\cite{M2300107}. }\label{tab:params}
\end{table}

For our numerical estimates, we mostly use the values of parameters planned for the future iteration ${\rm A}^\#$ of the LIGO GW detector \cite{M2300107}, see Table\,\ref{tab:params}. The only exception is the squeeze factor. In order to make our estimates more tolerant to the optical losses, we use the reduced value of this factor, 6\,dB instead of 10\,dB.  This relatively modest value corresponds to the reasonably optimistic value of the quantum efficiency of the interferometer, equal to
\begin{equation}
  \eta = \frac{1}{1+e^{-2r}} = 0.8 \,,
\end{equation}
see Ref.\,\cite{Demkowicz_PRA_88_041802_2013}.

Consider now two characteristic particular cases. The first one is the ``basic'' interferometric position meter, see \cite{12a1DaKh}. For this particular case, we suppose that the cross-corelation term in Eq.\,\eqref{S_gen} vanishes, which corresponds to the measurement of the phase quadrature of the output light. In this case, the ratio \eqref{xi2} is equal to
\begin{equation}\label{xi2_PM}
  \xi^2_{\rm PM}(\Omega) = \frac{1}{2}\biggl[
      \frac{e^{-2r}}{\mathcal{K}_{\rm PM}(\Omega)} + \mathcal{K}_{\rm PM}(\Omega)e^{2r}
    \biggr] ,
\end{equation}
where
\begin{equation}
  \mathcal{K}_{\rm PM}(\Omega) = \frac{\gamma\Theta_{\rm PM}}{\Omega^2(\gamma^2+\Omega^2)}
\end{equation}
is the position meter's optomechanical coupling factor \cite{02a1KiLeMaThVy},
$\gamma$ is the half-bandwidth of the interferometer \cite{Buonanno2003},
\begin{equation}\label{kTheta_PM}
  \Theta_{\rm PM} = \frac{8\omega_oI_c}{mcL}
\end{equation}
is the normalized optical power, $I_c$ is the total optical power circulating in the two arms of the interferometer, $\omega_o$ is the carrier light frequency,  $c$ is the speed of light, and $L$ is the arm(s) length. It is easy to see that the function \eqref{xi2_PM} touches its minimal value, equal to one and corresponding to the SQL, only at one given frequency, defined by the equation
\begin{equation}
  \mathcal{K}_{\rm PM}(\Omega)e^{2r} = 1 \,,
\end{equation}
and increases at both lower and higher frequencies.

Then, consider the speed meter, see Refs.\,\cite{12a1DaKh, 19a1DaKhMi}. Here we assume that a simple frequency-independent cross-correlation could be used: $S_{xF}=\const\ne0$. This type of the cross-correlation could be introduced by using a homodyne detector with some given homodyne angle $\zeta$. In this case, the factor \eqref{xi2} is equal to
\begin{equation}\label{xi2_SM}
  \xi^2_{\rm SM}(\Omega) = \frac{1}{2}\biggl[
      \frac{e^{-2r} + e^{2r}\cot^2\zeta}{\mathcal{K}_{\rm SM}(\Omega)}
      - 2e^{2r}\cot\zeta + \mathcal{K}_{\rm SM}(\Omega)e^{2r}
    \biggr] ,
\end{equation}
where $\mathcal{K}_{\rm SM}$ the speed meter's optomechanical coupling factor. Its exact form varies for different flavors of the speed meter. In the relevant to this paper cases of the ``canonical'' Sagnac speed meter, see Ref.\,\cite{Chen2002}, and of the double pass speed meter of Ref.\,\cite{18a1DaKnVoKhGrStHeHi}, it is equal to
\begin{equation}\label{K_SM}
  \mathcal{K}_{\rm SM}(\Omega) = \frac{\gamma\Theta_{\rm SM}}{(\gamma^2+\Omega^2)^2}
\end{equation}
where
\begin{equation}\label{Theta_SM}
  \Theta_{\rm SM} = \frac{16\omega_oI_c}{mcL} \,.
\end{equation}
The optical power $I_c$ here corresponds to the total optical power in the interferometer created by both passes of the light through it (in the scheme shown in Fig.\,\eqref{fig:scheme}, it is equal to the total optical power in both polarizations). The numerical factor in $\Theta_{\rm SM}$ is twice as big as in $\Theta_{\rm PM}$ because in the speed meter, the light interacts twice with the mechanical probe object.

It is easy to see that in the low-frequency band of $\Omega\ll\gamma$, the factor \eqref{K_SM} does not depend on $\Omega$:
\begin{equation}
  \mathcal{K}_{\rm SM}(\Omega) \to \const.
\end{equation}
It is this feature, common to all flavors of the speed meter, that allows optimization
of the speed meters sensitivity in a broad band without any additional elements, like the filter cavities.

In the case of $\zeta=\pi/2$ (the phase quadrature measurement), the function \eqref{xi2_SM} can be optimized by setting
\begin{equation}
  \frac{\Theta_{\rm SM}}{\gamma^3}e^{2r} = 1 \,.
\end{equation}
In this case, $\xi^2_{\rm SM}$ follows the SQL in the low frequency band:
\begin{equation}
  \xi^2_{\rm SM}(\Omega\ll\gamma) \approx 1 \,.
\end{equation}

The sensitivity exceeding the SQL in a broad band can be achieved by using the optimal value of the homodyne angle, defined by
\begin{equation}\label{ctz_opt}
  \cot\zeta = \frac{\Theta_{\rm SM}}{\gamma^3} \,.
\end{equation}
In this case,
\begin{equation}\label{xi2_LF}
  \xi^2_{\rm SM}(\Omega\ll\gamma) \approx \xi^2_{\rm LF}
  = \frac{\gamma^3}{2\Theta_{\rm SM}}e^{-2r} \,,
\end{equation}
where the subscript ``LF'' stand for the ``low-frequency''.

\begin{figure}
  \includegraphics[width=0.7\textwidth]{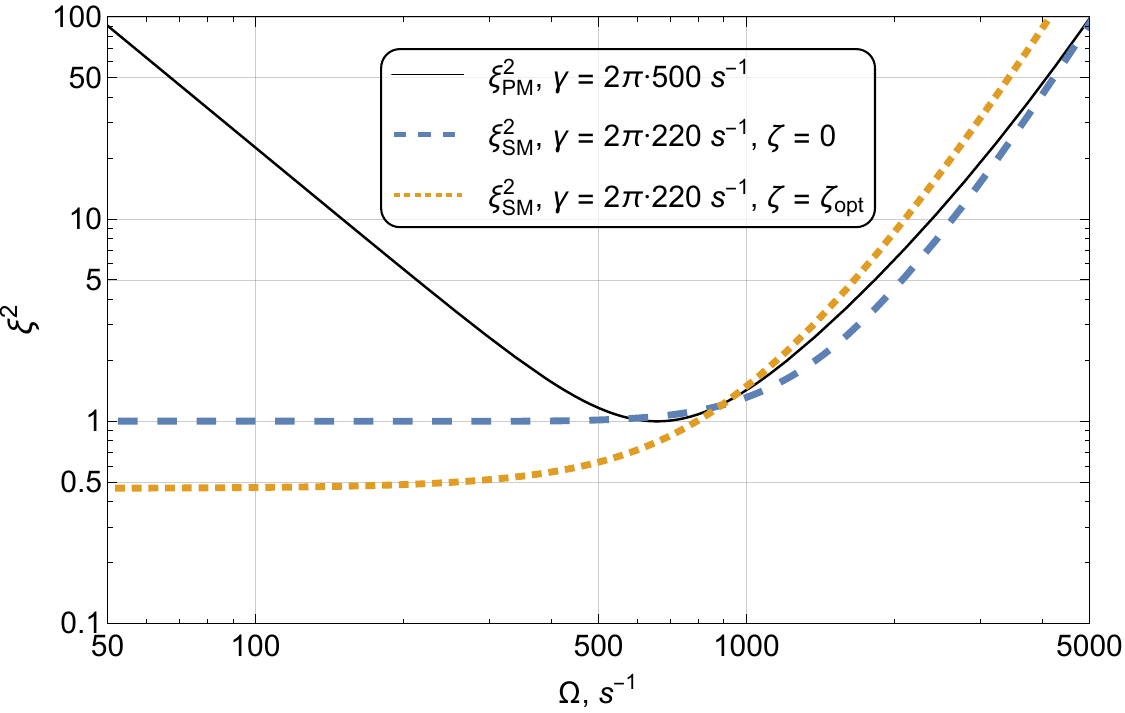}
  \caption{Examples of the factors $\xi^2_{\rm PM}$ and $\xi^2_{\rm SM}$.
  Thin solid: $\xi^2_{\rm PM}$, $\gamma=\pi\times500\,{\rm s}^{-1}$;
  long dashes: $\xi^2_{\rm SM}$, $\gamma=2\pi\times220\,{\rm s}^{-1}$, $\cot\zeta=0$;
  short dashes: $\xi^2_{\rm SM}$, $\gamma=2\pi\times220\,{\rm s}^{-1}$, $\cot\zeta$ is given by Eq.\,\eqref{ctz_opt}.
  All other parameters are listed in Table\,\ref{tab:params}.}\label{fig:intro}
\end{figure}

This consideration is illustrated by Fig.\,\ref{fig:intro}, where the functions \eqref{xi2_PM} and \eqref{xi2_SM} are plotted for some reasonable values of $\gamma$ and $\zeta$.

\section{Optimization of the quantum noise}\label{sec:S}

As shown in Appendix A, the spring factor of the scheme in  Fig.\,\ref{fig:scheme} is
\begin{equation}\label{K_gen}
  K(\Omega) = \frac{m\gamma\Theta_{\rm SM}}{4(\gamma-i\Omega)^2}\sin2\psi \,,
\end{equation}
and the component spectral densities that appear in Eq.\,\eqref{S_gen}, are, respectively, equal to
\begin{subequations}\label{S_comp_gen}
  \begin{gather}
    S_{xx}(\Omega) = \frac{\hbar}{2m\mathcal{K}_{\rm SM}(\Omega)}
      \frac{e^{2r}\cos^2\zeta + e^{-2r}\sin^2\zeta}
        {\gamma^2\cos^2\psi\cos^2\zeta + \Omega^2\sin^2\psi\sin^2\zeta} \,, \\
    S_{FF}(\Omega) = \frac{\hbar m\mathcal{K}_{\rm SM}(\Omega)}{2}
      (\gamma^2e^{-2r}\cos^2\psi + \Omega^2e^{2r}\sin^2\psi) \,, \\
    S_{xF}(\Omega) = -\frac{\hbar}{2}
        \frac{\gamma e^{-2r}\cos\psi\sin\zeta - i\Omega e^{2r}\sin\psi\cos\zeta}
          {\gamma\cos\psi\cos\zeta + i\Omega\sin\psi\sin\zeta} \,.
  \end{gather}
\end{subequations}
It is easy to see that in the particular case of $\psi=\pi/2$, the spring factor \eqref{K_gen} vanishes and Eqs.\,\eqref{S_comp_gen} reduce to the Sagnac interferometer case \eqref{xi2_SM}.

The rigorous optimization of $S_{\rm sum}$ have to be done numerically taking into account the technical (thermal etc) noise sources in the interferometer, as well as the shape of the signal. This task is beyond the scope of this article. Instead, we consider a simple universal procedure that gives quasi-optimal results and allows to demonstrate advantages of the combined "speed meter + optical spring" regime.

It is easy to see, that if the value of the angle $\psi$ is close to speed meter one $\pi/2$:
\begin{equation}\label{eps}
  \psi = \frac{\pi}{2} - \epsilon \,, \quad |\epsilon| \ll 1 \,,
\end{equation}
then in the considerable and important low-frequency band,
\begin{equation}
  \epsilon\gamma < \Omega < \gamma \,,
\end{equation}
Eqs.\,\eqref{S_comp_gen} take the form equal to the speed meter one. More specifically, keeping only the leading terms in $\epsilon$ in Eqs.\,\eqref{K_gen} and \eqref{S_comp_gen}, we obtain that they reduce to the following ones:
\begin{equation}\label{K_lf}
  K = m\Omega_0^2 \,,
\end{equation}
where
\begin{equation}
  \Omega_0 = \sqrt{\frac{\Theta_{\rm SM}\epsilon }{2\gamma}}
\end{equation}
is the mechanical eigenfrequency,
and
\begin{subequations}\label{S_xx_lf}
  \begin{gather}
    S_{xx}(\Omega) = \frac{\hbar}{2m\Omega^2}\sigma_{xx} \,, \\
    S_{FF}(\Omega) = \frac{\hbar m\Omega^2}{2}\sigma_{FF} \,, \\
    S_{xF}(\Omega) = \frac{\hbar}{2}\sigma_{xF} \,,
  \end{gather}
\end{subequations}
where
\begin{subequations}\label{sigmas}
  \begin{gather}
    \sigma_{xx} = \frac{\gamma^3}{\Theta_{\rm SM}}(e^{-2r} + e^{2r}\cot^2\zeta) \,, \\
    \sigma_{FF} = \frac{\Theta_{\rm SM}}{\gamma^3}e^{2r} \,, \\
    \sigma_{xF} =  e^{2r}\cot\zeta \,.
  \end{gather}
\end{subequations}
Note that the factors \eqref{sigmas} does not depend on $\Omega$.

Substituting Eqs.\,\eqref{K_lf}, \eqref{S_xx_lf} into Eqs.\,\eqref{xi2}, \eqref{S_gen} and taking into account that
\begin{equation}
  \sigma_{xx}\sigma_{FF} - \sigma_{xF}^2 = 1 \,,
\end{equation}
we obtain that the factor \eqref{xi2} can be presented as follows:
\begin{equation}\label{xi2_app}
  \xi^2(\Omega) = \frac{1}{2}\biggl[
      \biggl(\frac{\Omega_0^2}{\Omega^2} - 1 + \frac{\sigma_{xF}}{\sigma_{xx}}\biggr)^2
        \sigma_{xx}
      + \frac{1}{\sigma_{xx}}
    \biggr] .
\end{equation}

It is easy to see that the optical spring create a ``well'' in the quantum noise intensity at the frequency
\begin{equation}\label{Omega_min}
  \Omega_{\rm min} = \frac{\Omega_0}{\sqrt{1-\sigma_{xF}/\sigma_{xx}}} \,.
\end{equation}
and with the minimum, equal to
\begin{equation}\label{xi2_min}
  \xi^2_{\rm min} = \frac{1}{2\sigma_{xx}} \,,
\end{equation}
Above this frequency, but still in the area of $\Omega\ll\gamma$, the function \eqref{xi2_app} is approximately flat, similar to the ordinary speed meter case, and is equal to
\begin{equation}\label{xi2_shelf}
  \xi_{\rm shelf}^2 \approx \frac{1}{2}(\sigma_{XX} - 2\sigma_{Xf} + \sigma_{ff})
\end{equation}

The values of $\xi^2_{\rm min}$ and $\xi^2_{\rm shelf}$ depend on the homodyne angle $\zeta$ in different ways. Therefore, various approaches to optimization are possible, prioritizing either one or another. We assume that the factor $\xi^2_{\rm shelf}$ is more important, because it defines the broad-band low-frequency sensitivity, and minimize this quantity.

It is easy to show that the minimum of Eq.\,\eqref{xi2_shelf} in $\zeta$ is provided by Eq.\,\eqref{ctz_opt}, and the value of this minimum is equal to $\xi^2_{\rm LF}$, see Eq.\,\eqref{xi2_LF}. In this case, Eqs.\,\eqref{Omega_min} and \eqref{xi2_min} take the following forms:
\begin{gather}
  \Omega_{\rm min} = \Omega_0\sqrt{1 + \frac{1}{4\xi_{\rm SM}^4}} \,.\label{Omega_min_opt}\\
  \xi^2_{\rm min} = \frac{\xi^2_{\rm LF}}{1 + 4\xi^4_{\rm LF}} \,.\label{xi2_min_opt}
\end{gather}

As the final step of our quasi-optimization procedure, we substitute the value of $\zeta$ given by Eq.\,\eqref{ctz_opt}, into the rigorous equations \eqref{K_gen}, \eqref{S_comp_gen}, and then substitute the latter ones into Eqs.\,\eqref{xi2},\,\eqref{S_gen}. 

\begin{figure}
  \includegraphics[width=0.7\textwidth]{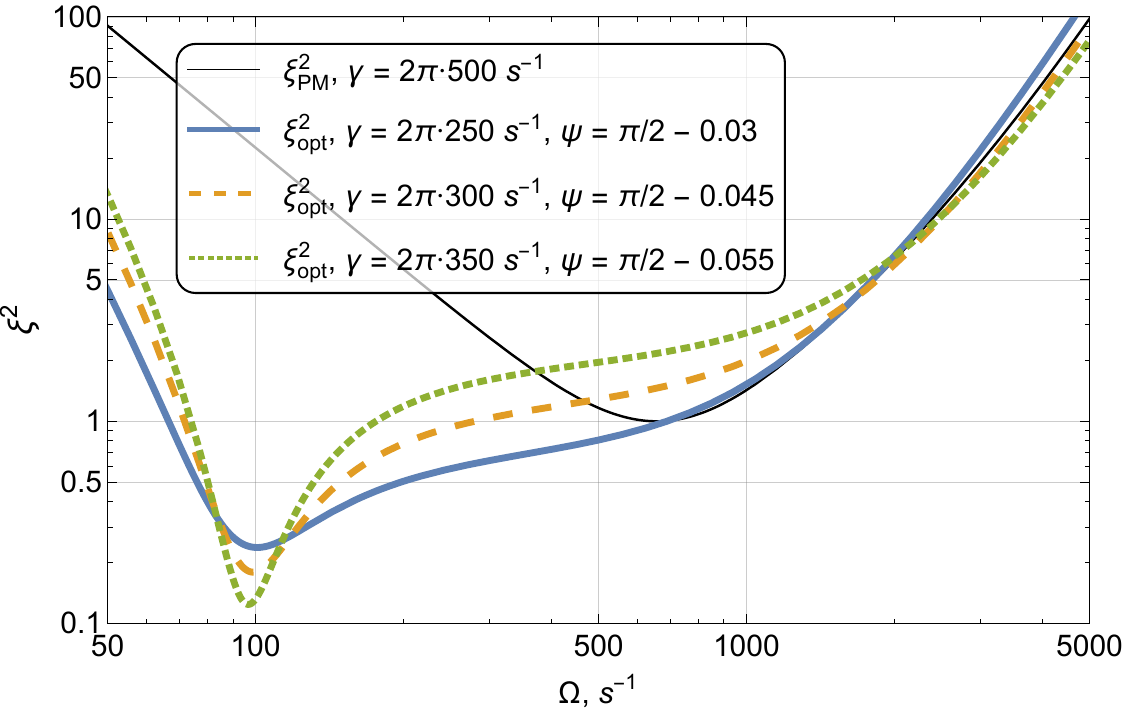}
  \caption{Plots of the factor $\xi^2_{\rm opt}$ for different values of $\Theta_{\rm SM}/\gamma^3$.
  Thick solid: $\gamma = 2\pi\times250{\rm s}^{-1}$, $\psi = \pi/2-0.03$;
  long dashes: $\gamma = 2\pi\times300{\rm s}^{-1}$, $\psi = \pi/2-0.045$;
  short dashes: $\gamma = 2\pi\times350{\rm s}^{-1}$, $\psi = \pi/2-0.055$.
  Thin solid: position meter, $\gamma = 2\pi\times500{\rm s}^{-1}$ (for the comparison).
  All other parameters are listed in Table\,\ref{tab:params}.
  }\label{fig:plots_1}
\end{figure}

\begin{figure}
  \includegraphics[width=0.7\textwidth]{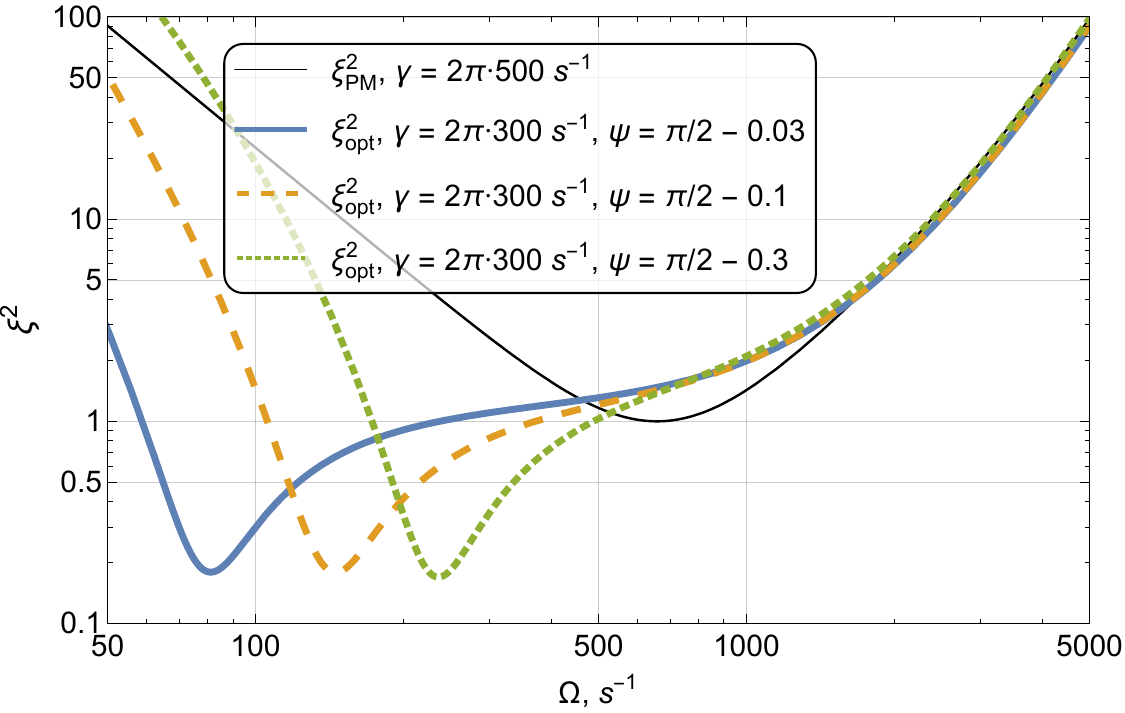}
  \caption{Plots of the factor $\xi^2_{\rm opt}$ for different values of $\psi$.
    Thick solid: $\gamma = 2\pi\times300{\rm s}^{-1}$, $\psi = \pi/2-0.03$;
    long dashes: $\gamma = 2\pi\times300{\rm s}^{-1}$, $\psi = \pi/2-0.1$;
    short dashes: $\gamma = 2\pi\times300{\rm s}^{-1}$, $\psi = \pi/2-0.3$;
    Thin solid: position meter, $\gamma = 2\pi\times500{\rm s}^{-1}$ (for the comparison).
    All other parameters are listed in Table\,\ref{tab:params}.
  }\label{fig:plots_2}
\end{figure}

The results are plotted in Figs.\,\ref{fig:plots_1} and \ref{fig:plots_2}. In Fig.\,\ref{fig:plots_1}, we vary the factor $\Theta_{\rm SM}/\gamma^3$ that appears in Eqs.\,\eqref{sigmas} in order to show how the shape of the quantum noise depends on this factor (actually, we vary $\gamma$, assuming that the value of $\Theta_{\rm SM}$ is given by Table \ref{tab:params}). It is easy to see from these plots, as well as from Eqs.\,\eqref{xi2_shelf}, \eqref{xi2_min_opt}, that $\xi^2_{\rm shelf}$ decreases with the increase of $\Theta_{\rm SM}/\gamma^3$, similar to the ordinary speed meter case, providing better broad-band sensitivity, but the value of $\xi^2_{\rm min}$ increases. As a result, the minimum in the quantum noise is pronounced only in the underpowered case of $\xi^2_{\rm shelf}\gtrsim1$ and dissolves in the very good sensitivity case of  $\xi^2_{\rm shelf}\ll1$.

In Figs.\,\ref{fig:plots_2}, we vary the phase shift $\psi$  in order to demonstrate, that the value of $\Omega_0$ can be tuned on demand. The trade-off is a degradation in sensitivity at frequencies below $\Omega_0$.

\section{Conclusion}\label{sec:conclusion}

In our opinion, the main advantages of the scheme considered here, which combines the double-pass quantum speed meter topology, proposed in Ref.\,\cite{18a1DaKnVoKhGrStHeHi}, with the double-pass optical spring concept proposed in Ref.\,\cite{24a1Kh}, are the following.

(i) In order to add the double-pass optical spring to the speedmeter scheme of \cite{18a1DaKnVoKhGrStHeHi}, it is necessary just to adjust the phase shift $2\psi$ between two passes of the light though the interferometer. Moreover, the original scheme of \cite{18a1DaKnVoKhGrStHeHi} also does not requires any radical changes in the  standard Michelson/Fabry-Perot interferometer topology of the contemporary GW detectors. It has to be emphasized, however, that the scheme \cite{18a1DaKnVoKhGrStHeHi}, and correpondingly, the one considered here, rely on the assumption that all parameters of the components of the interferometer (the mirrors and the beamsplitter) does not depend on the light polarisation.

(ii) The optical spring allows to increase the low-frequency sensitivity without increasing the optical power, see Ref.\,\cite{12a1DaKh}. It worth noting that the optical power in the contemporary GW detectors is barely enough to reach and slightly overcome the SQL, see Ref.\,\cite{Jia_Science_385_1318_2024}. Higher power is planned for future iterations of the detectors, but this increase will be offset by the approximately proportional increase in the mirror mass $m$, see \eg Ref.\,\cite{M2300107}.

(iii) The shape of the quantum noise of our scheme can be tuned flexibly also just by changing the phase shift $2\psi$, see Fig.\,\ref{fig:plots_2}. A very interesting option arises from this feature, namely the possibility of dynamically adjusting the frequency $\Omega_0$ in real time to follow the "chirp" GW signals, as it was proposed in  Refs.\,\cite{Meers_PRD_47_2184_1993, Simakov_PRD_90_102003_2014} and demonstrated using the table-top prototype scheme in Ref.\,\cite{Aronson_OL_49_6980_2024}.

\acknowledgments

This work was supported by the Theoretical Physics and Mathematics Advancement Foundation ``BASIS'' Grant \#23-1-1-39-1. The paper has been assigned LIGO document number P2500082.

The authors would like to thank D.\,Salykina and S.\,Danilishin for commenting on the manuscript.

\appendix

\section{Quantum noise of the two-pass interferometer}\label{app:S}

In this Appendix we extend calculations of Ref.\,\cite{24a1Kh} to the general case (without the bad cavity approximation). Following that work, we assume that the phase shift between the two interactions is introduced into the classical carrier fields. Evidently, this is equivalent to introducing the opposite phase shift into the quantum fields. At the same time, this approach simplifies the calculations.


Using the scaling law approach of Ref.\,\cite{Buonanno2003}, we model two passes of the probing light through the interferometer by the effective Fabri-Perot cavities, see Fig.\,3 of Ref.\,\cite{24a1Kh}. The input-output relations for these cavities are as follows, see Ref.\,\cite{12a1DaKh}:
\begin{equation}\label{io_12}
  \svector{\hat{b}_i^c}{\hat{b}_i^s} = \frac{\ell^*}{\ell}\svector{\hat{a}_i^c}{\hat{a}_i^s}
    + \frac{2k_oE}{\ell}\sqrt{\frac{\gamma}{\tau}}\svector{-\sin\phi_i}{\cos\phi_i}x  , \\
\end{equation}
where $i=1,2$ denotes the cavity number, $\hat{a}_i^c$, $\hat{a}_i^s$ are the two-photon quadratures of the incident fields, $\hat{b}_i^c$, $\hat{b}_i^s$ are the corresponding output quadratures,
\begin{equation}
  E = \sqrt{\frac{2I_{c1}}{\hbar\omega_o}}  = \sqrt{\frac{I_c}{\hbar\omega_o}}
\end{equation}
is the classical amplitude of the intracavity fields, $I_{c1}$ is the optical power circulating in each of the polarizations,
\begin{equation}
  \phi_{1,2} = \phi \pm \psi \,.
\end{equation}
are the phases of the intracavity fields, $k_o = \omega_o/c$, $\tau=L/c$, and
\begin{equation}
  \ell = \gamma - i\Omega \,.
\end{equation}
Taking into account that $\hat{a}_2^{c,s}=\hat{b}_1^{c,s}$, combining Eqs.\,\eqref{io_12}, and renaming
\begin{equation}
  \hat{a}_1^{c,s} \to \hat{a}^{c,s}\,, \quad \hat{b}_2^{c,s} \to \hat{b}^{c,s} \,,
\end{equation}
we obtain:
\begin{multline}\label{io_gen}
  \svector{\hat{b}^c}{\hat{b}^s} = \frac{\ell^*{}^2}{\ell^2}\svector{\hat{a}^c}{\hat{a}^s}
  + \frac{2k_oE}{\ell^2}\sqrt{\frac{\gamma}{\tau}}\biggl[
        \ell^*\svector{-\sin\phi_1}{\cos\phi_1} + \ell\svector{-\sin\phi_2}{\cos\phi_2}
      \biggr]\hat{x} \\
  = \frac{\ell^*{}^2}{\ell^2}\svector{\hat{a}^c}{\hat{a}^s}
    + \frac{4k_oE}{\ell^2}\sqrt{\frac{\gamma}{\tau}}\biggl[
          \gamma\svector{-\sin\phi}{\cos\phi}\cos\psi
          - i\Omega\svector{\cos\phi}{\sin\phi}\sin\psi
        \biggr]\hat{x} \,.
\end{multline}

The back action force acting on the probe object is equal to (see again Ref.\,\cite{12a1DaKh}):
\begin{equation}
  F_{\rm b.a.} = \frac{2\hbar k_oE}{\ell}\sqrt{\frac{\gamma}{\tau}}\sum_{i=1,2}
    (\cos\phi_i\ \sin\phi_i)\svector{\hat{a}_i^c}{\hat{a}_i^s}
  = \hat{F}_{\rm fl} - K\hat{X} \,.
\end{equation}
where
\begin{equation}
  \hat{F}_{\rm fl} = \frac{4\hbar k_oE}{\ell^2}\sqrt{\frac{\gamma}{\tau}}\bigl[
    (\gamma\cos\phi\cos\psi + i\Omega\sin\phi\sin\psi)\hat{a}^c
    + (\gamma\sin\phi\cos\psi - i\Omega\cos\phi\sin\psi)\hat{a}^s
    \bigr]  \label{f_fl_gen}
\end{equation}
and $K$ is goiven by Eq.\,\eqref{K_gen}.

Without loss of generality, we set
\begin{equation}
  \phi = -\frac{\pi}{2} \,.
\end{equation}
This choice yields equations that are consistent with those of the Sagnac interferometer. In this case, Eqs.\,\eqref{io_gen} and \eqref{f_fl_gen} simplify as follows:
\begin{gather}
  \svector{\hat{b}^c}{\hat{b}^s} = \frac{\ell^*{}^2}{\ell^2}\svector{\hat{a}^c}{\hat{a}^s}
    + \frac{4k_oE}{\ell^2}\sqrt{\frac{\gamma}{\tau}}
        \svector{\gamma\cos\psi}{i\Omega\sin\psi}\hat{x} \,, \\
  \hat{F}_{\rm fl} = -\frac{4\hbar k_oE}{\ell^2}\sqrt{\frac{\gamma}{\tau}}
    (i\Omega\hat{a}^c\sin\psi + \gamma\hat{a}^s\cos\psi) \,. \label{f_fl}
\end{gather}

The output signal of the homodyne detector is equal to
\begin{equation}
  \hat{b}_\zeta = \hat{b}^c\cos\zeta + \hat{b}^s\sin\zeta
  = \frac{4k_oE}{\ell^2}\sqrt{\frac{\gamma}{\tau}}
      (\gamma\cos\psi\cos\zeta + i\Omega\sin\psi\sin\zeta)(\hat{x}_{\rm fl} + \hat{x}) \,,
\end{equation}
where
\begin{equation}
  \hat{x}_{\rm fl} = \frac{\ell^*{}^2}{4k_oE}\sqrt{\frac{\tau}{\gamma}}
    \frac{\hat{a}^c\cos\zeta + \hat{a}^s\sin\zeta}
      {\gamma\cos\psi\cos\zeta + i\Omega\sin\psi\sin\zeta} \,.
\end{equation}

It can be shown using Eqs\,\eqref{S_SQZ} that spectral densities of $\hat{x}_{\rm fl}$ and $\hat{F}_{\rm fl}$, as well as their corresponding spectral density, are given by Eqs.\,\eqref{S_comp_gen}.


%

\end{document}